# An Improved Exact Algorithm for the Domatic Number Problem*

Tobias Riege, † Jörg Rothe, ‡ Holger Spakowski §
*Institut für Informatik*
*Heinrich-Heine-Universität Düsseldorf*
*Düsseldorf, Germany*

Masaki Yamamoto ¶
*Department of Mathematical and Computing Science*
*Tokyo Institute of Technology*
*Tokyo, Japan*

March 16, 2006

**Abstract**

The 3-domatic number problem asks whether a given graph can be partitioned into three dominating sets. We prove that this problem can be solved by a deterministic algorithm in time $2.695^n$ (up to polynomial factors). This result improves the previous bound of $2.8805^n$, which is due to Fomin, Grandoni, Pyatkin, and Stepanov. To prove our result, we combine an algorithm by Fomin et al. with Yamamoto's algorithm for the satisfiability problem. In addition, we show that the 3-domatic number problem can be solved for graphs $G$ with bounded maximum degree $\Delta(G)$ by a randomized algorithm, whose running time is better than the previous bound due to Riege and Rothe [RR05] whenever $\Delta(G) \geq 5$. Our new randomized algorithm employs Schöning's approach to constraint satisfaction problems.

*A two-page abstract of this paper is to appear in the Proceedings of the *Second IEEE International Conference on Information & Communication Technologies: From Theory to Applications*, April 2006, which improves on the results of a preliminary version [RR05].

†Email: riege@cs.uni-duesseldorf.de. Supported in part by the German Science Foundation under grants RO 1202/9-1 and RO 1202/9-3.

‡Corresponding author. Email: rothe@cs.uni-duesseldorf.de. Supported in part by the German Science Foundation under grants RO 1202/9-1 and RO 1202/9-3, and by the Alexander von Humboldt Foundation in the CONNECT program. Work done in part while visiting the Tokyo Institute of Technology.

§Email: spakowsk@cs.uni-duesseldorf.de. Supported in part by the German Science Foundation under grants RO 1202/9-1 and RO 1202/9-3.

¶Email: yamamot3@is.titech.ac.jp.

# 1  Introduction

A dominating set in an (undirected) graph $G$ is a subset $D$ of the vertex set $V$ of $G$ such that the closed neighborhood of $D$ equals $V$. The domatic number problem asks for a partition of $V$ into a maximum number of disjoint dominating sets. This number, denoted by $\delta(G)$, is called the domatic number of $G$. The domatic number problem arises in the area of computer networks and is related to the tasks of distributing resources and of locating facilities in the network. This problem has been intensely studied, see, e.g., [CH77, Far84, Bon85, KS94, HT98, FHK00, FHKS02, RR04].

For each $k \geq 3$, it is NP-complete to determine whether or not the domatic number of a given graph is at least $k$, see Garey and Johnson [GJ79]. That is why we cannot expect to find a polynomial-time algorithm that solves the problem.

Among the various ways of coping with NP-hard problems (such as approximation, randomization, or parameterized complexity), much attention has been paid to designing exact exponential-time algorithms for such problems that are better than the trivial exponential-time algorithm. In particular, if the trivial algorithm runs in time $3^n$ but one is able to find a $c^n$ algorithm for this problem with $c < 3$ (up to polynomial factors), then this algorithm can handle larger problem instances in the same amount of time than the trivial algorithm. This difference can be quite important in practice. For recent surveys on this subject, we refer to Schöning [Sch05] and Woeginger [Woe03].

The first result breaking the trivial $3^n$ barrier for the 3-domatic number problem is due to Riege and Rothe [RR05]. Fomin et al. [FGPS05] improved their result by providing a $2.8805^n$ bound. In this paper, we obtain a further improvement by making use of two known algorithms: an algorithm by Fomin et al. [FGPS05] for generating all minimal dominating sets of a graph, and Yamamoto's algorithm for the satisfiability problem [Yam05]. We show that these two algorithms can be combined so as to yield a $2.695^n$ time algorithm for the 3-domatic number problem.

In addition, we prove that there is a randomized algorithm solving this problem for graphs with bounded maximum degree, which improves the previous bound due to Riege and Rothe [RR05]. Here, we apply Schöning's results for constraint satisfaction problems [Sch99, Sch02], which previously was also useful in improving the bounds of randomized exponential-time algorithms for the satisfiability problem. For example, Iwama and Tamaki [IT04] designed a randomized algorithm with running time $\tilde{\mathcal{O}}(1.324^n)$ for solving 3-SAT by combining Schöning's algorithm with an algorithm due to Paturi et al. [PPSZ98].

# 2  Preliminaries

We first introduce some graph-theoretical notation. Graphs are pairs $G = (V, E)$, where $V$ is the vertex set of $G$ and $E$ is the edge set of $G$. All graphs considered in this paper are undirected and simple. That is, the edges of any graph are unordered pairs of vertices, and there are neither multiple nor reflexive edges; i.e., there exists at most one edge of the form $\{x, y\}$ for any two vertices $x$ and $y$, and there is no edge of the form $\{x, x\}$. Moreover, we require all graphs to not have isolated vertices. In general, graphs need not be connected.

**Definition 1** *Let $G = (V, E)$ be a graph.*

- *For any vertex $v \in V$, define the* neighborhood *of $v$ in $G$ by*

$$N(v) = \{u \in V \mid \{u, v\} \in E\},$$

  *and define the* closed neighborhood *of $v$ in $G$ by*

$$N[v] = \{v\} \cup N(v).$$

- *For any subset $U \subseteq V$ of the vertices of $G$, define $N[U] = \bigcup_{u \in U} N[u]$ and $N(U) = N[U] - U$.*

- *A subset $D \subseteq V$ is a* dominating set *of $G$ if and only if every vertex $u \notin D$ is a neighbor of some vertex $v \in D$. That is, $D$ dominates $G$ if and only if $N[D] = V$.*

- *A dominating set $D$ is called a* minimal dominating set *if and only if there exists no dominating set $C$ of $G$ with $C \subset D$.*

- *The* domatic number *of $G$ (denoted by $\delta(G)$) is the maximum number of disjoint dominating sets.*

- *Given a graph $G$ and a positive integer $k$, the $k$-domatic number problem,* k-DNP *for short, asks whether or not $\delta(G) \geq k$.*

Note that at least one partition of $G$ into $\delta(G)$ (i.e. the maximum number of) disjoint dominating sets contains $\delta(G) - 1$ minimal dominating sets.

For $k \geq 3$, the $k$-domatic number problem is NP-complete, see Garey and Johnson [GJ79]. Therefore, no polynomial-time algorithm for k-DNP exists unless P = NP.

Here, we focus on the case $k = 3$ only. The first algorithm beating the trivial $\tilde{\mathcal{O}}(3^n)$ barrier[1] for the 3-domatic number problem is due to Riege and Rothe [RR05]. They also investigate this problem for graphs with bounded maximum degree, and they propose a deterministic and a randomized algorithm for it. In particular, this deterministic algorithm outperforms their general deterministic algorithm for the 3-domatic number problem whenever $\Delta(G)$, the maximum degree of the input graph $G$, is in the range $3 \leq \Delta(G) < 7$.

Fomin et al. [FGPS05] recently improved the general result from [RR05] by constructing an algorithm for 3-DNP with running time $\tilde{\mathcal{O}}(2.8805^n)$. Their algorithm makes use of a new algorithm for enumerating all minimal dominating sets of a graph.

**Theorem 2 (Fomin et al. [FGPS05])** *There is an algorithm for listing all minimal dominating sets in an $n$ vertex graph $G$ in time $\tilde{\mathcal{O}}(1.7697^n)$.*

---

[1]As is common for exponential-time algorithms, we use the $\tilde{\mathcal{O}}$ notation to indicate that polynomial factors are neglected. That is, for functions $f$ and $g$, we write $f \in \tilde{\mathcal{O}}(g)$ if $f \in \mathcal{O}(p \cdot g)$, where $p$ is some polynomial.

Their proof relies on a new method to evaluate the size of the recursion tree for a given exponential-time algorithm. This technique, which is called "measure and conquer," was introduced by Fomin, Grandoni, and Kratsch [FGK05], who applied it to give a better analysis of the runtimes of exact backtracking algorithms for the minimum dominating set problem and the minimum set cover problem.

The "measure and conquer" technique is based on the following idea. By choosing a suitable measure of the subproblems generated by the recursive algorithm considered, one can lower-bound the progress made by the algorithm in each branching step. A clever choice of this measure can yield a much better worst-case runtime analysis of the problem, even though the algorithm considered is not new and has long been known.

Based on this new technique, Fomin et al. [FGPS05] design an exponential-time algorithm for determining the domatic number of a given graph (and thus, in particular, for solving 3-DNP). Their approach resembles the dynamic-programming algorithm by Lawler [Law76] for computing the chromatic number of a graph.

**Corollary 3 (Fomin et al. [FGPS05])** *There is an algorithm for computing the domatic number of a given graph that runs in time $\tilde{\mathcal{O}}(2.8805^n)$.*

Prior to this paper, the time bound stated in Corollary 3 was the best result known for the (search version of the) domatic number problem, and in particular for 3-DNP. In the next section, we further improve the result that follows from Corollary 3 for the decision problem 3-DNP.

## 3 An Improved Exact Algorithm for the 3-Domatic Number Problem

In this section, we improve the exponential running time for 3-DNP that follows from Corollary 3. To this end, we combine the algorithm for enumerating all minimal dominating sets from Theorem 2 with an algorithm due to Yamamoto [Yam05].

Yamamoto's algorithm, which is based on (and improves) an algorithm due to Hirsch [Hir00], solves the NP-complete problem SAT in time $\tilde{\mathcal{O}}(1.234^m)$, where $m$ is the number of clauses of the given boolean formula in conjunctive normal form. To prove Theorem 5 below, we will apply Yamamoto's algorithm to a special version of SAT, which is called NAE-SAT ("not-all-equal satisfiability").

**Definition 4** *Let $\varphi = \varphi(X, C)$ be a boolean formula in conjunctive normal form consisting of a collection $C = \{c_1, c_2, \ldots, c_m\}$ of $m$ clauses over the variable set $X$. We say $\varphi$ is in NAE-SAT if and only if there exists a truth assignment for $X$ satisfying all clauses in $C$ and such that in none of the clauses, all literals are true.*

**Theorem 5** *There exists an exact algorithm solving the 3-DNP problem in time $\tilde{\mathcal{O}}(2.695^n)$.*

**Proof.** We will only sketch the proof.

Let $G = (V, E)$ be a given graph with $n$ vertices. Using the algorithm from Theorem 2, generate all minimal dominating sets $D \subseteq V$ of $G$. Given one such minimal dominating set $D$, create a formula $\varphi_D = \varphi_D(X, C)$ for the NP-complete problem NAE-SAT as follows:

- The set of variables is defined as
$$X = \{x_v \mid v \in V\}.$$

- For each vertex $v \in V$, create the clause
$$C_v = \left\{ \bigcup_{\substack{u \in N[v] \\ u \notin D}} x_u \right\},$$
so the clause set is defined as $C = \{C_v \mid v \in V\}$.

Note that $G \in$ 3-DNP if and only if $\varphi_D$ is in NAE-SAT for some formula $\varphi_D$ thus defined.

(Alternatively, one might formalize this 3-DNP instance $G$ with some minimal dominating set $D$ as an instance $(\mathcal{U}, \mathcal{S}_D)$ of the minimum set cover problem, where $\mathcal{U} = \{0, 1, 2\}$ is a universe of elements (which correspond to the three dominating sets of $G$ if $G$ is in 3-DNP), and where $\mathcal{S}_D$ is a correspondingly defined nonempty set system over $\mathcal{U}$. Then, we assign the element 2 of $\mathcal{U}$ to $D$, and $G \in$ 3-DNP if and only if every set $S$ in $\mathcal{S}_D$ has both the values 0 and 1. This corresponds exactly to the NAE-SAT property stated above.)

The number of clauses in $\varphi_D$ equals $n$, the number of vertices in $G$. Using the standard polynomial-time reduction from NAE-SAT to SAT (see Garey and Johnson [GJ79]), which adds the negation of each clause to the clause set, we obtain a formula $\varphi'_D = (X, C \cup \overline{C})$ with the property that
$$\varphi_D \in \text{NAE-SAT} \iff \varphi'_D \in \text{SAT}.$$

Using the exponential-time SAT algorithm designed by Yamamoto [Yam05], we can now determine the satisfiability of $\varphi'_D$ in time $\tilde{\mathcal{O}}(1.234^{2n})$. Note here that $\varphi'_D$ has $2n$ clauses. It follows that our algorithm to solve 3-DNP has a running time of
$$\tilde{\mathcal{O}}(1.7697^n \cdot 1.234^{2n}) = \tilde{\mathcal{O}}(2.695^n),$$

which completes the proof. ∎

## 4 An Improved Randomized Algorithm for the 3-Domatic Number Problem

We now turn to the case where the maximum degree $\Delta(G)$ of the input graph $G$ is bounded by some small constant. Riege and Rothe [RR05] described a randomized algorithm for this problem. Here, we observe that their result can be improved by using Schöning's algorithm for constraint satisfaction problems [Sch99, Sch02].

| $\Delta(G)$ | 3 | 4 | 5 | 6 | 7 | 8 | source |
|---|---|---|---|---|---|---|---|
| deterministic | $2.2894^n$ | $2.6591^n$ | $2.8252^n$ | $2.9058^n$ | $2.9473^n$ | $2.9697^n$ | [RR05] |
| randomized | $2^n$ | $2.3570^n$ | $2.5820^n$ | $2.7262^n$ | $2.8197^n$ | $2.8808^n$ | [RR05] |
| randomized | $2.2501^n$ | $2.4001^n$ | $2.5001^n$ | $2.5715^n$ | $2.6251^n$ | $2.6667^n$ | Thm. 6 |

Table 1: Three 3-DNP algorithms for graphs $G$ with bounded maximum degree $\Delta(G)$.

**Theorem 6** *There is a randomized algorithm solving the 3-DNP problem for graphs $G$ with bounded maximum degree $\Delta(G)$, whose running time depends on $\Delta(G)$ as stated in Table 1.*

**Proof.** The 3-DNP problem for a graph $G$ can easily be formulated as a constraint satisfaction problem (CSP) on the domain $D = \{0, 1, 2\}$ with each constraint having order $\ell = \Delta(G) + 1$. The order of a constraint is the number of arguments in the constraint.

Let $G = (V, E)$ be a given graph. Create the constraint satisfaction problem $F$ with $n$ variables as follows

- The set of variables is defined as

$$X = \{x_v \mid v \in V\}.$$

- For each vertex $v \in V$, create the constraint $C_v$ defined by

$$C_v(x_v, x_{w_1}, x_{w_2}, \ldots, x_{w_{\|N[v]\|-1}}) = 1$$
$$\Longleftrightarrow$$
all values of $D = \{0, 1, 2\}$ appear in the values of $x_v, x_{w_1}, x_{w_2}, \ldots, x_{w_{\|N[v]\|-1}}$,

where $w_1, w_2, \ldots, w_{\|N[v]\|-1}$ are the vertices adjacent to $v$.

It is easy to see that

$$G \in \text{3-DNP} \iff \text{CSP } F \text{ has a solution}.$$

To determine whether $F$ has a solution, we can apply Schöning's randomized algorithm for solving constraint satisfaction problems [Sch99, Sch02].

We thus obtain a randomized algorithm for 3-DNP with running time

$$\tilde{\mathcal{O}}\left(3\left(1 - \frac{1}{\Delta(G) + 1}\right) + \epsilon\right)^n,$$

for any $\epsilon > 0$. ∎

For graphs with $\Delta(G) \geq 5$, this is an improvement over the randomized algorithm given by Riege and Rothe [RR05], see Table 1.

# 5 Conclusions

In this paper, we considered the 3-domatic number problem, which asks whether a given graph can be partitioned into three dominating sets, and we have shown how to improve on existing exact and randomized exponential-time algorithms for this problem.

In particular, by reducing 3-DNP to the NAE-SAT problem and by combining Yamamoto's algorithm [Yam05] with the algorithm by Fomin et al. [FGPS05], we obtained an exact (i.e., deterministic) algorithm that runs in time $\tilde{\mathcal{O}}(2.695^n)$. This result improves on the previously best bound of $\tilde{\mathcal{O}}(2.8805^n)$ for the 3-domatic number problem, which was achieved by Fomin et al. [FGPS05]. An even earlier—and to our knowledge, the first—nontrivial bound for this problem is due to Riege and Rothe [RR05] who presented an $\tilde{\mathcal{O}}(2.9416^n)$ algorithm.

We also described a randomized algorithm solving the 3-domatic number problem for graphs $G$ with bounded maximum degree $\Delta(G)$. Whenever $\Delta(G) \geq 5$, the running time of our new randomized algorithm is better than the previously known bound, which is due to Riege and Rothe [RR05]. These results are summarized in Table 1. Our new randomized algorithm makes use of Schöning's algorithm for constraint satisfaction problems [Sch99, Sch02].

**Acknowledgment.** We are grateful to Osamu Watanabe and Dieter Kratsch for inspiring discussions on the subject of this paper. In particular, we thank Dieter Kratsch for calling our attention to Lawler's algorithm. The second author thanks Osamu Watanabe for kindly hosting his research visit at the Tokyo Institute of Technology in September and October 2005 that brought about this collaboration.